# Digital Twins in Biopharmaceutical Manufacturing:
# Review and Perspective on Human-Machine Collaborative Intelligence

Mohammed Aatif Shahab[a*], Francesco Destro[a], Richard D. Braatz[a*]

[a]Department of Chemical Engineering,
Massachusetts Institute of Technology, Cambridge, MA, 02139, United States

*Corresponding author's email: braatz@mit.edu, moha0095@mit.edu

**Abstract**

The biopharmaceutical industry is increasingly developing digital twins to digitalize and automate the manufacturing process in response to the growing market demands. However, this shift presents significant challenges for human operators, as the complexity and volume of information can overwhelm their ability to manage the process effectively. These issues are compounded when digital twins are designed without considering interaction and collaboration with operators, who are responsible for monitoring processes and assessing situations, particularly during abnormalities. Our review of current trends in biopharma digital twin development reveals a predominant focus on technology and often overlooks the critical role of human operators. To bridge this gap, this article proposes a collaborative intelligence framework that emphasizes the integration of operators with digital twins. Approaches to system design that can enhance operator trust and human-machine interface usability are presented. Moreover, innovative training programs for preparing operators to understand and utilize digital twins are discussed. The framework outlined in this article aims to enhance collaboration between operators and digital twins effectively by using their full capabilities to boost resilience and productivity in biopharmaceutical manufacturing.

# 1    Introduction

The biopharmaceutical industry is rapidly expanding its manufacturing capabilities in response to growing demand and anticipation of unexpected periods of high need, such as the COVID-19 pandemic [1]. To address these expectations, there is a strong need to design manufacturing processes that increase throughput while ensuring product quality. Regulatory initiatives such as quality-by-design [2] and process analytical technology (PAT) [3] have been promoting this modernization effort to enhance the quality and productivity of pharmaceutical processes [4], [5]. Emerging technologies such as continuous processing, process analytical technology (PAT), digitalization, active process control, and process modeling are pivotal tools to achieve this purpose [3],[6],[7],[8],[9]. To scale-up biopharmaceutical processes to high throughput while ensuring the product quality, it is crucial to manage the propagation of impurities and disturbances. Deviations from normal operating conditions can alter quality attributes, critical process parameters, and the performance of equipment, posing risks to patient safety and leading to non-compliance with current good manufacturing practices (cGMP) standards [10]. On average, each process deviation in the (bio)pharmaceutical industry costs between $25,000 to $55,000, with potential financial impacts that can exceed $1,000,000 in cases of product loss [1].

Statistics indicate that human error accounts for the majority of deviations from normal operating conditions [11]. For instance, a survey conducted by BioPhorum indicated that human error accounts for about 50% of deviations in biopharmaceutical processes [11]. Human errors are not only frequent but also extremely costly [12]. In the incident at Emergent BioSolutions, human error led to the destruction of 15 million vaccine doses, costing the company $180 million and resulting in a 37% drop in share price [13]. Beside the economic damage, deviations from normal operating conditions can also result in safety-related incidents involving thermal runaways and explosions, which can cause project delays and safety risks [14]. To address these challenges, biopharmaceutical companies are increasingly developing digital twins – virtual replicas of physical processes [5], [9], [15], [16]. Digital twins play a crucial role in managing the propagation of impurities and disturbances by predicting future process states and enabling timely interventions to maintain drug quality. Through real-time simulations, sophisticated control algorithms, and process analytics, digital twins enable automated monitoring and control of manufacturing

processes. However, while automation can reduce some types of errors, it cannot address all the complexities of manufacturing. Incidents related to human error still persist across manufacturing industries that already adopted advanced automation [17], [18]. Further, automated systems and digital twins still face several limitations, such as data misinterpretation, incorrect system configurations, and unforeseen anomalies that require human oversight and intervention [19], [20]. Simply removing humans from the process does not eliminate the risk of errors; in fact, it can create new risks that automation alone cannot mitigate. In a highly automated system, operators may be required to intervene only when the system fails, which can happen at the worst possible times, when the capabilities of the automated control system of the plant have already been overcome. Additionally, these rare interventions can create confusion about whether a fault arises from a system malfunction or a malicious cyber intrusion, which complicates crisis response [21], [22]. This occasional involvement during a critical scenario can overwhelm even skilled operators, who are required to act during pivotal system failures.

This paper argues that, while digital twins are valuable for enhancing biopharmaceutical manufacturing, their full potential can only be realized through collaborative intelligence that incorporates human expertise. Collaborative intelligence emphasizes the interaction of humans and machines by using their respective abilities to enhance decision-making and process control. While automation excels at performing repetitive and predictable activities, human intelligence is essential for handling unexpected abnormalities, complex decision-making, and interpreting nuanced data [23]. Biopharma industries can become resilient, robust, and effective by promoting cooperation between human operators and digital twins [24].

In this paper, we first review the literature on digital twin technologies within the biopharmaceutical sector. Although recent reviews discuss digital twins in biopharmaceutical manufacturing [25],[26], many use the term "digital twin" so loosely that industry expectations and actual functionalities often fail to align. This inconsistent usage also complicates standardization and cross-comparison among different frameworks, making it harder to evaluate digital twin solutions in practice. In addition, existing articles rarely include a clear framework for differentiating the developmental stages of digital twins. Moreover, current studies adopt a technology-centric perspective and overlook the human expertise essential for effective design and operation. In this paper, we systematically organize digital twin approaches around an established definition, and explain how process modeling, PAT, process monitoring, and advanced control each contribute to digital twin development across the various stages of biopharmaceutical manufacturing. We then provide a perspective on how industries can effectively integrate collaborative intelligence by incorporating human expertise into digital twin development and operation. This approach will ultimately lead to improved decision-making, adherence to safety standards, and optimal performance in biopharmaceutical manufacturing.

## 2  The state of digital twin implementation in biopharmaceutical manufacturing

## 2.1  Foundations of digital twin technology

The concept of digital twin was first introduced by Michael Grieves in 2002, initially referred to as the "Ideal for product lifecycle management [27],[28], which refers to the process of managing the entire lifecycle of a product from its initial design and development to its final disposal or recycling. In 2010, NASA's John Vickers formally introduced the phrase "digital twin" [29]. A digital twin comprises three main components: a physical entity that either exists or is intended to exist, its virtual or digital counterpart, and the communication channel that allows data and information to flow between the two [30],[31]. This bidirectional communication ensures that data from the physical world can be fed into the virtual model, while insights and information from the virtual model can influence the physical counterpart. Notably, a physical twin is not required for a digital twin to be genuine: the defining characteristic of a digital twin is its intention to be a digital representation that will eventually correspond to a physical entity [31]. Digital twins are classified into three types based on their developmental stage and intended function: digital twin prototype (DTP), digital twin instance (DTI), and digital twin aggregate (DTA) [30],[31]. The DTP is created at the early stages of product development, before the physical entity exists. A DTP depicts a product prototype, which can include several possible design variations, and functions as a virtual model for design, simulation, optimization, and control. DTPs can aid in the detection and correction of possible issues early in the design process. A DTI represents the real-time digital equivalent of an existing physical system. DTIs enable continuous monitoring, simulation, optimization, and control of the physical system using data collected from sensors. A DTA is a collection of DTIs linked together to allow real-time data sharing and analysis across multiple units or processes.

These categories can be naturally extended to biopharmaceutical digital twins. We therefore employ the same DTP, DTI, and DTA framework to categorize current literature and industrial implementations in this field. A summary of digital twin types and their benefits in biopharmaceutical manufacturing is shown in Table 1. A typical biopharmaceutical process begins with upstream processing, which involves production of a biological substance of interest (active pharmaceutical ingredient, API). This step is followed by downstream processing, which involves the purification of the API from process- and product-related impurities. The final step is the formulation of the final drug product to be administered to the patient. Each of these processing stages comprises multiple unit operations. Developing models to simulate the behavior of biopharmaceutical unit operations without the existence of a physical process falls under the category of DTP, which can use mechanistic, data-driven, or hybrid strategies to simulate and optimize control strategies. Once the physical process is established, engineers integrate sensors for real-time measurement and data transfer between physical and digital instances, transitioning to a DTI stage. PAT tools are key enablers of DTIs. A DTA of a biopharmaceutical process is finally established when DTIs across unit operations in both upstream and downstream processing are connected for plant-wide monitoring and control. The following section will discuss recent developments in process modeling that contribute to the advancement of digital twins in biopharmaceutical manufacturing.

## 2.2 Process Modeling

The foundation of any digital twin lies in the accurate replication of biophysiochemical processes through mathematical modeling. This section explores various modeling approaches used in digital twin development: mechanistic modeling (based on fundamental physical and chemical principles), data-driven modeling (leveraging historical data and machine learning to predict process dynamics), and hybrid modeling (combining mechanistic and data-driven approaches). Table 2 shows benefits and limitation of each of these modeling approaches and Table 3 summarizes recent work on process modelling in biopharmaceutical manufacturing, most of which falls under DTP category.

### 2.2.1 Mechanistic models

Mechanistic models form the foundation of digital twin technology by providing insights into biological processes and their dynamics. Monoclonal antibodies (mAbs) are currently the predominant modality in the biopharmaceutical industry, with over 100 new mAbs entering the development phase each year and market size expected to reach $300 billion by 2025 [32],[33], [34]. mAbs are biological drugs that can target specific antigens found on cancer cells, pathogens,

or in inflammatory pathways, thus allowing for personalized medical interventions with high efficacy and specificity. An important aspect of mAb production is the optimization of mAb glycosylation, the biochemical process through which sugar groups are added to proteins. Glycosylation patterns significantly affect the efficacy, stability, and safety of mAbs [35]. Several mechanistic models for mAb manufacturing have been developed to optimize operating conditions for higher titer and process efficiency [36],[37],[38],[39],[40]. Recent advancements include models that predict the effect of feeding strategies on cell metabolism and glycan profiles, with the goal of optimizing nutrient conditions to enhance mAb glycosylation efficiency [37],[39]. Other studies have focused on developing models that explain how feeding strategies affect the concentration of nucleotide sugar donors which influence the attachment of glycans to antibodies during glycosylation process [40].

Mechanistic modeling has also played a crucial role in the optimization of mRNA manufacturing, with a special focus on the in vitro transcription (IVT) process in upstream mRNA production [41],[42],[43],[44]. Akama et al. developed a mechanistic model of the IVT process that focuses on the interaction between RNA synthesis and the formation of magnesium pyrophosphate, a byproduct that can reduce the reaction rate [43]. More recently, Stover et al. developed the first mechanistic model for IVT that describes the kinetics of nucleation and growth of magnesium pyrophosphate crystals, which are known to have a detrimental effect on the IVT reaction [44]. The model was validated with experimental data and quantitatively predicted previously unexplained phenomena, such as the role of pyrophosphatase in enhancing IVT productivity by degrading pyrophosphate crystals.

Mechanistic modeling is also a pivotal instrument for enhancing the manufacturing process of novel modalities, such as viral vectors for gene therapy [45]. Recombinant adeno-associated virus (rAAV), the primary vector for commercial in vivo gene therapies, can provide a cure against genetic disorders and other severe diseases by delivering therapeutic genes to a patient's cells [46]. However, current manufacturing processes for rAAV are inefficient and cannot meet the rising demand of vectors for rAAV-based gene therapy [47],[48]. These inefficiencies also result in high rAAV manufacturing costs, which can exceed $300,000 per dose. Recently, the first mechanistic models for rAAV production through the main processes used in commercial-scale rAAV manufacturing have been developed, including for mammalian [49] and insect [50] cell platforms. The models have been used to identify and tackle the process and genetic bottlenecks of current manufacturing processes [51]. Recent work has extended the models to optimize continuous rAAV manufacturing in mammalian [52] and insect [53] cells, demonstrating that continuous rAAV production can lead to a significant increase of productivity and reduction of manufacturing cost.

Current mechanistic models for downstream biopharmaceutical processes have a primary focus on chromatography for removal of product- and process-related impurities [54],[55],[56],[57]. Monoclonal antibodies are purified by passing them through a column filled with resin, a material that binds specific molecules based on properties such as size, charge, or affinity [58]. The process begins with Protein A chromatography, which captures mAbs and removes process-related impurities like host cell proteins, followed by ion-exchange chromatography (IEX) to eliminate fragments, aggregates, and charge variants [59],[60]. The choice of resin and its binding capacity, along with conditions such as pH and salt concentration, impacts the separation efficiency. Consequently, various studies have focused on developing mechanistic models to predict mAb purification efficiency by optimizing resin properties (such as type of resins and their binding capacity) and operating conditions [54],[61],[62],[63]. These models also help assess the effectiveness of new resins in removing impurities from mAbs [64] and have the potential to reduce the need for experimental work by over 75% [65]. Researchers have also developed models for multimodal chromatography, a technique that combines different binding mechanisms (such as charge and hydrophobic interactions) to enhance purification [66],[67],[68]. Recently, Hess et al. developed a mechanistic model for multimodal chromatography that predicts binding mechanism of various mAbs under different pH conditions [68]. These models have been used in digital twins with real-time data transfer with a physical plant, achieving iterative model updates and automatic process control [57].

Mechanistic models for optimizing the lyophilization of biopharmaceutical processes are also reported in the literature [69],[70],[71],[72]. Lyophilization, or freeze-drying, is a critical step for maintaining stability and extending the shelf life of biopharmaceuticals. This process involves three main steps: freezing, primary drying, and secondary drying [70]. First, the product is frozen to turn most of the water into ice. Then, during primary drying, the vials containing the product are placed on shelves inside a chamber to remove ice through sublimation [73]. Finally, any remaining water bound to the product is removed by heating in a secondary drying step. Primary drying is the longest and most energy-demanding step and optimization of this step is crucial [74]. Key parameters such as shelf temperature and chamber pressure directly influence the sublimation rate. Uneven drying, particularly in vials located at the edges or corners of the chamber, further complicates the process [71]. Consequently, mechanistic models have been developed to optimize process parameters for the primary drying phase, with a particular focus on accurately representing heat transfer through the vial side walls [69],[70]. More recently, Srisuma et al. developed the first mechanistic model to predict drying times during the primary drying step for all vials—including inner, edge, and corner vials—across traditional freeze-drying, microwave-assisted freeze-drying, and hybrid freeze-drying [71]. The model has also been extended to support continuous lyophilization processes [72].

Mechanistic models are invaluable tools for the development of digital twins. However, data-driven models can enhance prediction accuracy in certain instances, as outlined in the next section.

### 2.2.2 Data-driven models

Data-driven models based on data analytics and machine learning have contributed to the development of digital twins with high predictive accuracy in biopharmaceutical manufacturing [75],[76],[77],[78],[79]. Data-driven models have been applied in upstream bioprocessing for predicting key parameters (e.g., biomass and protein concentrations), optimizing feeding strategies, and finding optimal conditions for culture media in microbial and mammalian cell bioreactors [75],[80]. For instance, data-driven models based on artificial neural networks (ANNs) have been used to predict biomass and recombinant protein concentrations in *E. coli* fed-batch fermentation [81]. Seber and Braatz. demonstrated the use of ANNs to predict glycan distribution at N-glycosylation sites in CHO cells [78]. The ANN model was trained on two datasets: one showing the effects of gene knockouts for enzymes involved in glycan synthesis and another capturing the impact of changes in nucleotide sugar donor concentrations over time. The ANN models achieved a median prediction error of 9.10% in predicting glycan profiles, significantly outperforming previous models. These models suffered from overfitting due to inadequate separation of training, validation, and test data, and a lack of optimization in hyperparameter settings [78]. Data-driven models have also found several applications in downstream biopharmaceutical manufacturing, including in the optimization of filtration processes, in the selection of chromatography resins, and in the prediction of filter capacity and fouling rates during protein purification [75], [82], [83]. For instance, data-driven models were utilized to optimize column-sizing strategies of the chromatography process for purification of antigen-binding fragment (Fab) products [82].

While both mechanistic and data-driven models are useful individually, combining these approaches into hybrid models can lead to even more accurate models.

### 2.2.3 Hybrid models

Hybrid models combine the strengths of mechanistic and data-driven approaches to improve the accuracy of model predictions. A detailed review on the application of hybrid models in biopharmaceutical manufacturing can be found in Sokolov et al. [84] and Narayanan et al. [85]. Several studies have focused on developing hybrid models to support mAb manufacturing [86], [87]. For instance, in the fed-batch production of mAbs in CHO cells, a hybrid model was developed by combining ANNs with mechanistic mass balance equations to predict the time evolution of product titer, viable cell density, osmolality, and metabolites [86]. The model demonstrated superior accuracy and robustness compared to data-driven models when tested on a dataset of 81 fed-batch runs. Figure 1 illustrates a case study involving the prediction of lactate profiles in fed-batch mAb runs. Purely data-driven partial least squares (PLS) models diverged significantly from experimental measurements, whereas a hybrid model remained closely aligned with experimental data across the entire time span. Moreover, the hybrid model demonstrated superior extrapolation performance compared to data-driven PLS models when trained on 'low-titer' runs (<580 mg/L) and tested on 'high-titer' runs (>580 mg/L). The scaled RMSE of the hybrid model increased by only 5–10%, whereas the purely data-driven PLS models exhibited a much larger increase of 50–80% [86]. Similarly, Yatipanthalawa et al. [87] developed and validated hybrid models for predicting key performance metrics in the fed-batch production of mAbs in CHO cells. The study compared three different hybrid models, each combining the same mechanistic model with a different type of machine learning model: ANN, random forest, and XGBoost. No significant differences were observed in prediction capability among the three hybrid models, although the model exploiting the ANN in the data-driven compartment required a significantly larger computational time.

Although process models have been deployed in several biopharmaceutical processes, most literature studies fall into the category of DTP, since real-time data exchange with physical equipment was not achieved. In biopharmaceutical manufacturing, PAT tools can provide real-time process measurements and are key enablers for the development of digital twins, as discussed in the next section.

## 2.3 Process Analytical Technology (PAT)

Real-time measurement of process parameters is essential to transition from DTPs to DTIs in biopharmaceutical manufacturing. Within the QbD framework, critical process parameters (CPPs) and critical quality attributes (CQAs) should be monitored and controlled to guarantee product quality [88],[89]. In traditional (bio)pharmaceutical manufacturing, CPPs and CQAs are typically measured through offline sampling, which can take several hours or even days to generate results [90]. Recent advancements in the implementation of PAT in biopharmaceutical manufacturing are making possible to measure in real-time several CPPs and CQAs in both upstream and downstream production [91],[92], as summarized in Table 4. Researchers have employed PAT tools to measure key process parameters such as pH, dissolved oxygen, product related impurities, metabolite concentrations, product titer, protein aggregation, viable cell density, and glycosylation profiles [93]. Accurate real-time measurement of these parameters is crucial for maintaining optimal bioreactor conditions and product quality. For this, advanced techniques such as Raman spectroscopy, UV-visible spectroscopy, fluorescence spectroscopy, and vibrational spectroscopy are employed. These techniques are often coupled with chemometric methods—which use mathematical and statistical techniques to interpret complex data—and machine learning models to improve the prediction accuracy of process parameters [94]. Recent developments in PAT for mAb manufacturing include the use of Raman spectroscopy to identify the effect of thermal and oxidative stress on secondary and tertiary structures of mAbs [95]. Additionally, Raman spectroscopy has been combined with convolutional neural networks to monitor concentration of different forms of monoclonal antibodies that have slight charge differences (called charge variants) during cation exchange chromatography [96].

Soft sensors can be used for monitoring CPPs and CQAs for which real-time measurements are either not available or too noisy, by exploiting process data correlated to the variable of interest [97], [98]. During development, soft sensors are trained to capture the relationships between measured process variables and the unmeasured attributes of interest through mathematical modeling. Applications of soft sensors in biopharmaceutical manufacturing include the estimation of cell and metabolite concentrations in mammalian cell cultures using measurements from fluorescence and Raman spectroscopy [99],[100]. Additionally, soft sensors have been developed to monitor residual ice during primary drying and residual moisture during secondary drying in the lyophilization process [101],[102]. For instance, Fig. 2 illustrates a state observer based on a mechanistic model for secondary drying. The observer merges the model predictions with real-time temperature measurements to generate accurate estimates of unmeasured variables, even when the data are noisy or incomplete.

Real-time measurements obtained from PAT tools can be seamlessly transmitted from physical systems to digital twins using communication protocols like open platform communication (OPC) and its unified architecture variant (OPC-UA) [23], [103], as shown in Fig. 3. A review of the state of the art in communication protocols between physical equipment and digital twins can be found in Profanter et al. [104]. Many systems are now integrated with cloud-based platforms, which offer real-time data access and analysis to operators for monitoring and managing processes remotely [105]. Cloud infrastructures ensure that digital twins remain up to date with live process data and support scalability to improve decision-making.

While PAT provides real-time measurement data, this information must be mapped to either normal operating conditions or deviations. Process monitoring plays a crucial role in this mapping and is therefore a key enabler of DTIs, as discussed in the next section.

## 2.4 Process Monitoring in Biopharmaceutical Manufacturing

Effective process monitoring is critical for ensuring the quality of drugs. When manufacturing processes deviate from normal operating conditions, fault detection and diagnosis algorithms are essential to promptly alert operators and guide them in resolving incidents [106]. In industrial practice, fault detection and diagnosis algorithms predominantly rely on data-driven models [107], [108], especially multivariate statistical approaches. Among these, principal component analysis (PCA) and other latent variable models have emerged as cornerstone techniques, offering robust anomaly detection capabilities in complex pharmaceutical manufacturing environments [109], [110], [111]. While mechanistic models can be employed for fault detection and diagnosis, they often yield higher false alarm and missed fault rates compared to data-driven models, primarily due to process-model mismatch. As a result, the main application of mechanistic models in process monitoring lies within state estimation frameworks, where they serve as soft sensors for monitoring CQAs and CPPs for which real-time measurements are not available or too noisy [112], [113], [114]. Recently, hybrid models combining mechanistic and data-driven approaches have emerged as a promising solution for fault detection and diagnosis in (bio)pharmaceutical processes [115], [116]. Hybrid monitoring algorithms have demonstrated superior detection and diagnosis performance compared to state-of-the-art purely data-driven or mechanistic models [115] [116]. Recently, a hybrid model was utilized for process monitoring of a perfusion process in mAb production [23].

## 2.5 Process Control in Biopharmaceutical Manufacturing

Advanced process control (APC) plays a vital role in advancing digital twin technology by providing real-time feedback and making adjustments to the physical process. APC can significantly improve process yield and robustness, particularly in the presence of unexpected disturbances [118]. Product quality has been historically controlled at open-loop in pharmaceutical manufacturing. However, closed-loop control of accessory equipment and certain critical process parameters is regularly implemented in biopharmaceutical plants, such as for the temperature regulation of upstream cultures [119]. FDA described a three-level control structure for improving (bio)pharmaceutical quality [6]. Level 3, the current state of the art, focuses on end-product testing with limited process flexibility. In the past decade, there has been significant effort to move toward more advanced control systems, such as Level 2 and Level 1. Level 2 allows for process flexibility by defining a design space—a multivariate region of process parameters and critical material attributes within which CQAs are guaranteed to be in control. A Level 2 control strategy reduces reliance on end-product testing. Level 1 emphasizes real-time monitoring and automatic adjustments, enabling real-time product release. While this is the best way to ensure product quality, it has not yet been widely implemented in industrial practice. APC systems like model predictive control (MPC) and supervisory control and data acquisition (SCADA) play a key role in implementing Levels 1 and 2.

Accordingly, studies have developed MPC strategies to control upstream and downstream manufacturing processes through mechanistic models as well as data-driven models [118] [120] [121]. A summary of recent applications of APC in biopharmaceutical manufacturing can be found in Table 5. These studies primarily focus on automating the control of upstream processes, such as mAb production, and downstream processes, such as controlling chromatography. For instance, in fed-batch cell culture processes, a mechanistic-model-based MPC was developed to regulate titer and glycosylation profiles by adjusting process variables such as agitation speed, dissolved oxygen, and feed rates [120]. Another study used machine-learning-based MPC to optimize feeding strategies, maximizing cell growth and metabolite production, while keeping all process variables within desired limits [121]. The MPC system demonstrated better performance than a rule-based control technique. Several studies have focused on improving downstream process control in biopharmaceutical manufacturing [57],[122],[123],[124]. For instance, control strategies to manage the switching of column loading using feedforward control have been developed to enhance product yield [122], while acoustic wave separation systems with deep neural networks and distributed control systems have been used to improve CHO cell clarification and manage process deviations [123],[124]. APC has also been implemented in continuous viral inactivation [125]. For instance, Hong et al. developed a model-based control system for a column-based continuous viral inactivation process [125]. The controller regulates CPPs like pH and minimum residence time by optimizing feed flow rate using a pH controller and real-time monitoring of residence time distribution. A SCADA system has also been implemented to control an integrated continuous platform for mAb manufacturing that encompassed various unit operations such as a perfusion bioreactor, counter-current chromatography, virus inactivation, and two polishing steps [126].

As discussed previously, process modeling, PAT, process monitoring, and advanced control methods collectively enable the development of reliable digital twins in biopharmaceutical manufacturing. By continuously feeding PAT data into a state estimator (the core of the digital twin), unmeasured or noisy variables can be inferred in real time [102]. The digital twin uses live data to update process models and simulate future trajectories [57], detect impending faults, and optimize critical process and optimize scheduling in the unit operations [126]. A real-time digital twin requires robust computation; for example, [57] demonstrate that parallel computing and solver acceleration can enhance digital twin performance by lowering computational times. However, even more aggressive optimization may be needed to satisfy strict scheduling windows so that the next processing step is not delayed. In doing so, the digital twin closes the loop between real-time measurement, modeling, and control.

Overall, the current state of biopharmaceutical digital twin implementation primarily falls into the categories of DTPs, with a few studies falling into the DTI and DTA category (Table 1, Table 3–5). These studies tend to be highly technology-focused, emphasizing model development, measurement, and control, as shown in Fig. 3. Next, we discuss how a collaborative intelligence framework can be systematically implemented in the development of biopharmaceutical digital twins.

## 3 Human-machine Collaborative Intelligence in Biopharmaceutical Manufacturing

While digital twins are tools meant to support human decision-making by providing actionable insights [31], the current studies often overlook the crucial role of human operators who interact with the digital twin, as depicted in Fig. 3. Human operators remain crucial in monitoring process parameters, handling deviations, and ensuring compliance with cGMP, especially during abnormal process operations. However, without integrating human factors, digital twins risk being underutilized or misapplied by the very people they are designed to assist [127], [128]. To maximize their potential, digital twins need to work alongside human intelligence, supporting active involvement in decision-making and promoting a "collaboration over automation" approach in biomanufacturing environment. Organizations can enhance productivity when operators and automated systems collaborate effectively. Therefore, promoting collaboration between humans and machines is of critical importance, especially in biopharmaceutical manufacturing, where strict quality norms have to be adhered to. The human-machine collaborative intelligence framework presented in this work, as shown in Fig. 4, emphasizes the importance of designing digital twins that are intuitive and user-friendly, and of preparing operators to use these systems effectively. A well-designed digital twin, combined with trained operators, ensures that both entities – automation system and operators – can work together seamlessly (Fig. 4). However, for this collaboration to be truly effective, there must be a foundation of trust. The next section discusses how digital twins can be designed to build operator's trust.

## 3.1 Enhancing Trust in Automation

Trust in automation and digital twins is important for the effective implementation of advanced control systems in biopharmaceutical manufacturing. Trust ensures that operators are willing to rely on automated systems which are essential for optimizing processes. Trust encompasses multiple dimensions, including initial use, misuse, disuse, and abuse of automation [129]. These dimensions are highly relevant in the biopharmaceutical sector for deploying advanced process control (APC) algorithms, such as MPC. Operators may misuse APC by over-relying on it, disuse APC due to distrust (often stemming from false alarms or lack of understanding) or abuse APC, by deploying it without fully considering the consequences. APC can be non-intuitive for operators accustomed to single-loop control systems, which can lead to operators disabling APC when they do not fully understand or trust its control actions [130],[131],[132],[133]. As shown in Fig. 5, if the automation capabilities exceed trust, operator disuses it while, if trust exceeds the automation capabilities, the operator misuses it [128]. Therefore, proper calibration of trust is critical to ensure high-quality product and operation of biopharma processes [134].

Trust in automation depends on the operator's understanding of how context affects the system's capabilities, organizational and cultural factors, and transparency, which ensures that the automation's capabilities are clearly conveyed to the user [128]. For instance, transparency allows operators to understand the decision-making processes of the system [135]. For automation systems to be transparent, they must provide clarity and predictability in their actions. Operators are more likely to accept and follow the recommendations of an automated system if they can see and understand how these recommendations were derived [135],[136].

Transparency in the underlying components of digital twins can lead to better decision-making by operators [137]. In a study involving the implementation of machine learning models for lead-time prediction – the total time taken from the initiation of a process to completion — in biopharmaceutical quality control, a laboratory engineer emphasized the importance of understanding and trusting the model's results [138]. Building operator's trust in the model relied on understanding its predictions, inputs, and lead time calculations, as well as discussing its components. Bhakte et al. proposed using explainable artificial intelligence based on the Shapley framework to identify and explain the importance of different process variables in a fault detection model [139]. This approach reveals which variables contribute most to a fault, allowing operators to understand why a model makes certain decisions, and thus can enable them to make more informed operational choices in real-time [140]. In a study on diagnosing the faults in lithium-ion batteries (sensor failure, overheating, and short-circuiting), it was found that simply explaining a model's predictions to the operators may not be enough to build trust [141]. Operators frequently identified mismatch between the model predictions and the actual battery conditions, which ultimately reduced their trust. This issue may stem from operators' unfamiliarity with the limitations of the model. Future research should explore extending the timeframe of user interaction to assess whether trust increases as operators gain a better understanding of the model's behavior and performance.

Despite the growing importance of explainable and transparent algorithms, their implementation in real-world biopharmaceutical settings faces several challenges. Limited computational resources, reliance on proprietary 'black-box' vendor solutions, and the need to comply with stringent cGMP guidelines all hinder widespread adoption. Furthermore, tailoring algorithms to accommodate varying levels of operator expertise or acceptance often requires significant investments of time and resources, complicating efforts to standardize trust and validation measures across multiple sites.

Even so, transparency has tangible benefits [142]. For instance, in a predictive maintenance task, participants were provided baseline (no explanation), normative ("why" the system recommends a choice), or normative and contrastive ("why" and "why not") explanations for automated decisions. Decision times dropped significantly with the inclusion of explanations, while an increase in trust in automation occurred under explanatory conditions (Fig. 6). These findings confirm that operators who understand the rationale of automated decision-making become more confident and decisive in following automated suggestions.

The distribution of tasks between humans and automation systems is another critical element in building trust. It is important to design collaborative human-automation systems with clear delineations of "who does what with what information" [143]. For instance, while the automation system might alert the operator and suggest control actions for a potential issue, the operator can then use their expertise to determine the best course of action. In biopharmaceutical manufacturing, if the system detects unexpected cell growth rates in a bioreactor, the operator may decide to adjust nutrient feeds or change bioreactor conditions. This collaboration ensures that the strengths of both humans and automation are utilized effectively. Such a clear role definition ensures that operators are not overwhelmed with unnecessary information, thereby preventing mental overload and "mode confusion", which can happen when operators struggle to understand the current state of the system and their role within it [144].

Achieving human trust in the automation system involves multiple facets, including proper training, ensuring transparency in the system's operations, and clearly defining the roles of human operators and automation (Fig. 4). A significant contribution to human trust in machines comes from the design of the human-machine interface (HMI), as discussed in the next section.

### 3.2 Design of Human Machine Interface

The HMI is the source of all information exchange between digital twin, physical process, and operator. A well-designed HMI can enhance transparency by presenting information in a clear and accessible manner, making the system's decision-making processes understandable to the operators. A user-centric HMI enables operators to accomplish their duties effectively and with minimum errors [16]. During continuous process operation, particularly in the startup and shutdown phases, operators often switch between manual and automatic operating mode. This dynamic interaction requires a robust HMI that can support quick responses to disturbances and abnormal situations. Sand and Terwiesch emphasize that enhancing the performance of HMIs is crucial for operators who supervise automated operations and manage disturbances [145]. An effective HMI must be capable of providing clear, actionable information that supports the operator's decision-making processes, since actionable information improved trust and reduced decision time (Fig. 6). For developing a user-centric HMI, it is necessary to involve end users in the design process [23],[138]. During the development of a machine learning model for lead time prediction in biopharmaceutical quality control, user feedback was crucial for enhancing the usability of the graphical user interface (GUI) of the system. In a case study in biopharmaceutical manufacturing, Staff responsible for production planning reported that the first version of the GUI was non-intuitive and required a manual for its use. Prototyping allowed to develop an intuitive GUI, which improved the operators learning curve for using the machine learning model for lead time prediction [138].

A well-designed HMI should use visual elements effectively to communicate complex ideas and information clearly and quickly. For example, trends and predictive displays can improve the operator's ability to foresee potential issues and take preemptive actions [16],[146]. Trends help operators predict future states, thus enhancing situational awareness (SA) and supporting better decision-making. Consistency in design is crucial for usability. Consistency requires that dialog syntax (language, color, size, location, etc.) and semantics (behaviors associated with objects) are coherent throughout the interface. For instance, alarm colors such as bright red and yellow should be reserved exclusively for alarm conditions to avoid confusion. Inconsistent colors will make it harder for operators to interpret information correctly [147]. To further enhance HMI design, it is essential to build an information hierarchy and group related information together. An effective information hierarchy helps the operator to create a process overview and easily locate the needed level of details. Grouping related information together helps the operator perceive important connections, thus facilitating the comprehension of the information shown on the display [148]. Shahab et al. demonstrated that operator performance during interaction with a chemical process was enhanced when related information was grouped together on the HMI [149]. The study evaluated operator performance using two different HMI designs. In the first design, process variables were displayed based on their physical location in the plant (e.g., tray temperature of top plates displayed at the top of the distillation column). In the second design, the HMI was restructured to group related information together based on how operators use the data (e.g., all tray temperatures were grouped to help operators detect disturbances more easily). The second design resulted in enhanced operator performance.

Studies have also developed adaptive HMIs that adjust based on the operator's expertise, current task and level of stress, providing the right amount of information at the right time. To achieve this balance, it is important to analyze how operators interact with the system, including their behavioral patterns and responses to different process conditions, and then tune the HMI accordingly [144]. For example, during a procedural task such as setting up a machine for a specific job, the HMI can be adapted based on the operator's expertise [150]. For expert operators, the HMI might present streamlined information, allowing them to quickly select tools by inputting tool codes, while for less experienced operators, the HMI can display visual aids and step-by-step instructions to guide them through the process.

While a user-centric design of HMIs is fundamental for effective human-machine interaction, it alone does not guarantee that operators will not make mistakes. However, adapting HMIs to different operator profiles (varying cognitive styles, job roles, experience levels) requires significant up-front design and testing. Real-time personalization methods, such as adaptive layouts that change with operator workload, remain largely experimental and demand frequent software updates, on-site customization, and rigorous validation. As a result, most current industrial interfaces remain one-size-fits-all which limits the practical rollout of advanced HMI concepts in regulated environments. For this reason, most companies train all operators on HMI usage and advanced system interactions, as discussed in the next section.

## 3.3 Workforce Training

As the biopharma industry shifts towards digitalized and automated operation, it requires a workforce equipped with new skill sets [151]. Operators serve as the primary end users of digital twins, and without adequate preparation, it is not possible to fully achieve a successful collaboration. However, a techno-centric approach to digital twins can inadvertently reduce operators' understanding of these systems. In a recent survey, 67% of the respondents declared that the top two obstacles to implementing digitalization in biopharma manufacturing are the lack of skills and perceived complexity [152]. The lack of digital skills is a significant impediment to increasing digitalization and modeling in the sector. Staff retirement, turnover, shortage of skilled workers, and ineffective knowledge retention during role handovers further complicate this issue [153],[154],[155]. Regulatory agencies worldwide emphasize the importance of proper training in their cGMP guidelines. However, these guidelines often lack specific details on how training should be conducted [156], highlighting the urgent need for more effective training programs to bridge the skill gap and ease the adoption of digital twins in biopharmaceutical manufacturing.

Trained operators who are knowledgeable about digital twin systems can effectively collaborate with these technologies [157]. Nonetheless, the industry has struggled to keep up with rapid technological advancements in training methodologies, often relying on outdated practices [156]. Conventional training methods, such as reading standard operating procedures (SOPs) [158], are often perceived as tedious and ineffective, leading to poor compliance and increased errors [159]. According to a survey by the Parenteral Drug Association (PDA), an average pharmaceutical company manages around 1,250 cGMP-mandated SOPs, with the document management workload accounting for approximately 10–15% of total operating costs [154],[159]. A large-scale biopharmaceutical company spends 3.5 million hours annually on reading/understanding SOPs, with only around 10% of the acquired knowledge being effectively retained [158]. Current training procedures are generally neither clear, concise, accurate, nor user-friendly, and the systems managing them are poorly designed and maintained. Moreover, procedures are often released just in time for use, leaving little opportunity for comprehensive training. This training approach is both time-consuming and mentally exhausting, often leading to non-compliance with regulations and productivity loss. The current training schemes treat training as a mere 'tick-the-box' exercise, rather than effectively conveying the skills needed to perform tasks competently. This approach frequently leads to inadequate and inconsistent technical training, since SOPs are not designed to include all the information necessary for individuals to effectively learn and execute their duties [159]. While on-the-job training is typically more effective, it is also resource-intensive and expensive, with industries investing an estimated $7 billion annually in employee training [160].

Interactive training that includes real-life scenarios can be particularly beneficial to overcome the limitations of current training programs. For instance, operators in biopharmaceutical manufacturing can be trained on scenarios such as responding to equipment malfunctions, managing deviations in the bioreactor environment, or handling contamination events during the production process. Training with simulators can enable operators to practice abnormal scenarios in a controlled setting, enhancing their readiness and confidence in managing actual situations [161]. For example, operators can use digital twins to simulate the occurrence of a temperature spike in a bioreactor, and test different strategies to bring the temperature back to safe levels. Contamination events can also be simulated, to practice the identification of the best containment and corrective measures without the risk of actual product loss or contamination spread.

Adding immersion to these interactive simulations by using virtual reality (VR) can significantly improve training outcomes [162],[163]. Immersive VR provides an engaging and interactive learning environment, allowing employees to practice as in real life, but without disrupting ongoing production. Recently, a training program used VR to simulate the precise procedures required for pH calibration and adjustment in a biopharmaceutical plant [164]. After VR-based training, the trainees' practical skills were assessed by performing pH calibration with real equipment, with a metrology expert evaluating their performance. The study found that VR training was as effective as real-life training for teaching practical skills [164]. A recent study at the National Horizons Center (United Kingdom) deployed a VR-based penicillin production simulator (Fig. 7), featuring a 20 L steam-in-place bioreactor [165]. The simulator lets participants practice tasks like pH probe calibration and responds to faults such as sudden vessel-pressure spikes or incorrect feed-rate adjustments. In total, 40 engineering students and bioscientists reported strong engagement with these immersive scenarios. While VR and simulator-based training can accelerate learning, these programs are expensive to create and maintain. Many operators also lack foundational data-analytics skills, which compels companies to fund upskilling or risk underusing digital tools. One solution involves structuring training around a competency matrix, which has been shown to reduce the total theoretical course load for pharmaceutical operators by approximately 80%, from 137 hours to just 55 hours [166].

An effective training program includes not only the delivery of content but also an evaluation of its effectiveness [163]. Traditional assessment methods, such as questionnaires and decision-based evaluations, often fall short due to their subjective nature [167]. Simulator-based measures, which assess process behavior during operator interaction, and operator behavioral metrics, including actions performed and response times, offer more objective insights [168]. However, these methods do not explain the rationale behind an operator's actions. Given the complexity of biopharmaceutical processes, it is crucial to assess whether operators have accurate mental models, namely a correct internal understanding of the system. Cognitive systems engineering addresses this need by focusing on the alignment between an operator's mental models and the actual system [163], thereby enhancing decision-making and process performance.

### 3.4 Cognitive systems engineering

With humans playing a crucial role in interacting with digital twins in biopharmaceutical manufacturing, it is essential to understand the collaborative dynamics between the digital twin and operators. While the operation of digital twins through data, models and algorithms is well-documented, operators process information through complex cognitive functions. Cognitive systems engineering (CSE) aims to understand human cognitive processes and their interaction with digital twins by examining, for instance, how operators develop mental models of a process, make decisions under pressure, and respond to changes and anomalies [169]. Understanding how operators interpret digital twin outputs ensures that interface designs and data visualizations support effective decision-making. By aligning system design with human cognition, CSE helps to create HMI and training programs that promote more robust human–machine interactions, forming a beneficial feedback loop (Fig. 8). A recent study reported that, when dealing with deviations from normal operating conditions in a process, operators go through several mental states, such as identifying disturbances, setting goals, generating and confirming hypotheses, and rejecting incorrect hypotheses when outcomes do not match expectations [170]. A flaw in any step of this mental process can propagate into the physical process, potentially leading to significant errors. Hence, it is necessary to understand cognitive processes of operators to enhance decision-making and mitigate errors.

Traditionally, cognitive processes of operators have been assessed through subjective techniques, such as the NASA Task Load Index (NASA-TLX) [171] and the Subjective Workload Assessment Technique (SWAT) [172]. However, these methods are often criticized for their subjectivity and inability to provide real-time assessment, limiting their utility in digitalized operations [173]. With advancements in sensor technology, human cognitive processes can now be tracked using objective measurements from physiological sensors such as eye tracking and electroencephalogram (EEG). Eye tracking can record eye movement of operators when they interact with digital twins via HMI, while EEG can monitor their cognitive workload in real time. Studies have demonstrated that eye tracking can pinpoint how operators interact with the industrial processes [170],[174],[175]. For example, eye tracking has revealed that process operators struggle to effectively monitor process trends over time, indicating that they do not fully understand the effect of control actions, which can result in delays and inefficiencies in process control [174]. Eye tracking is also an invaluable tool for assessing training outcomes. A study reported that operators increasingly focused their eye gaze on the most relevant information sources on HMIs as they learned the causal relationships within the system, indicating improved understanding and performance [176]. Similarly, eye tracking studies comparing novice and expert operators revealed different cognitive strategies, with experts paying more attention to critical areas of the HMI than novices [177],[178],[179]. These insights can be used to deliver personalized feedback for better learning and training. Furthermore, understanding operator cognition through eye tracking can help design intuitive HMIs [149],[180],[181] for the biopharmaceutical digital twins. For instance, measures such as fixation-to-importance ratio (FIR) and selective attention effectiveness (SAE) were developed to evaluate effectiveness of information sources on the HMI [180]. Low FIR and SAE values indicate poor attention to important attention, prompting a change in the display of information.

Cognitive workload is another crucial aspect of operator interaction with automation that can be evaluated using physiological sensors [182]. Cognitive workload refers to the mental effort required to perform a task and is a critical determinant of operator performance [173]. When the cognitive workload exceeds the operator's working memory capacity, it leads to suboptimal performance and increased reliance on automation. EEG is an objective tool for assessing cognitive workload. Iqbal et al. used EEG to evaluate cognitive workload during training of operators dealing with disturbances in a chemical process [183]. Cognitive measures identified in the study detected a decrease in cognitive workload as operators developed a better understanding of the process. In particular, power spectral density of specific EEG frequency bands can indicate a mismatch between an operator's expectations and actual process behavior, signaling increased cognitive workload and potential failures in controlling process abnormalities [173].

Converting cognitive metrics such as eye gaze data or EEG signals into actionable interface adjustments or operator alerts remains challenging. Real-time collection and analysis require robust IT infrastructure and specialized algorithms.

## 4   Conclusions

This article discussed recent progress in the development and application of digital twins for biopharmaceutical manufacturing, highlighting the critical need for effective collaboration between human operators and digital twins. Significant advancements have been made in process models – mechanistic, data-driven, and hybrid – to create in silico replica of physical biopharmaceutical processes. However, most literature contributions focus on DTPs, used for simulating and optimizing processes before physical implementation. Recent developments in PAT are enabling the development of DTIs, which achieve real-time data transfer between physical system and digital twin, with applications in advanced process monitoring and control. A few studies have progressed to DTAs, integrating multiple DTIs across upstream and downstream manufacturing. Nonetheless, the development of digital twins remains highly techno-centered in the state of the art, often neglecting the role of operators that interact with digital twins. To address this issue, this article presented a collaborative intelligence approach that exploits the intelligence of human operators and the capabilities of digital twins for a synergistic enhancement of biopharmaceutical processes. Strategies have been discussed to achieve effective human-machine collaboration by enhancing operator trust in digital twins and HMI usability. Advanced training methods have been presented to improve operators' understanding of manufacturing processes and digital twins by using simulators and virtual reality. Effective training, coupled with performance assessment, can prepare skilled operators for next-generation biopharmaceutical manufacturing. Considering operator cognitive dynamics is also crucial for the design of digital twins. Techniques such as eye tracking and electroencephalography allow to monitor operators' cognitive activity while using digital twins. This information can be used to optimize the design of the digital system to ensure effective human-machine collaboration.

Looking ahead, future research should move beyond pilot projects and deploy digital twins across fully integrated, continuous manufacturing lines to capture end-to-end process dynamics. Future work should also explore the evolutionary relationship between operators and digital twins, with focus on how long-term interaction shapes trust, enhances productivity, and accelerates workforce skill development.

**Acknowledgment**

This research was supported by the U.S. Food and Drug Administration under the FDA BAA-22-00123 program, Award Number 75F40122C00200.

**Table 1: Summary of digital twin types and their benefits in biopharmaceutical manufacturing.**

| Digital twin classification | Typical use | Benefits in biopharma | Challenges | References |
|---|---|---|---|---|
| DTP (Digital twin prototype) | • Virtual replica before physical system exists<br>• Used in early design, simulation, and optimization | • Accelerates process development by allowing in silico experimentation<br>• Reduces the number of costly wet-lab trials<br>• Aids in design-space exploration and technology transfer | • May not capture all biological complexities if experimental data are limited<br>• May be subject to model mismatch as physical system evolves | [39, 40, 43, 44, 49-51, 53, 61, 63, 64, 68-71, 78, 81, 82, 86, 87] |
| DTI (Digital twin instance) | • Real-time digital counterpart of an existing physical system<br>• Updated continuously with live sensor data | • Enables real-time monitoring & control of CPPs/CQAs<br>• Facilitates proactive fault detection & diagnosis<br>• Supports closed-loop control | • Requires validated models and compliance with cGMP<br>• Models have to be regularly updated | [57,122-125] |
| DTA (Digital twin aggregate) | • Union of multiple DTIs across a production line or entire plant | • Improves scheduling, resource allocation, and integrated continuous manufacturing<br>• May reduce overall production costs and time to market | • Data management/ interoperability becomes more complex at plant scale<br>• Synchronizing multiple DTIs can be challenging when they differ in modeling depth or update frequency<br>• Regulatory compliance spans across multiple unit operations | [126] |

**Table 2: Advantages and disadvantages of mechanistic, data-driven, and hybrid modeling in biopharmaceutical manufacturing.**

| Approach | Key advantages | Key disadvantages | References |
|---|---|---|---|
| Mechanistic | <ul><li>Physically interpretable (rooted in first principles)</li><li>Better extrapolation performance</li><li>Can facilitate QbD and process optimization</li></ul> | <ul><li>Model development can be lengthy and requires domain expertise</li><li>Sensitive to parameter uncertainty and mismatch</li><li>Can be challenging to fully capture biological complexity</li></ul> | [36-44, 49-56, 61-72] |
| Data-driven | <ul><li>Rapid model building if sufficient data is available</li><li>Captures high-dimensional correlations</li><li>Less reliance on domain knowledge</li></ul> | <ul><li>"Black box" nature limits interpretability and trust</li><li>Overfitting risk</li><li>Extrapolation beyond training domain is unreliable</li></ul> | [76-83] |
| Hybrid | <ul><li>Combines mechanistic interpretability with data-driven flexibility</li><li>Yields more robust, accurate predictions</li><li>Can reduce total experimental effort</li></ul> | <ul><li>Requires both domain and machine learning expertise</li><li>Integrating two modeling paradigms can raise computational overhead</li></ul> | [84-87] |

**Table 3: Current trends in modeling biopharmaceutical processes. DTP = digital twin prototype, DTI = digital twin instance, DTA = digital twin aggregate**

| Type of model | Process stage | Digital twin category | Key contributions and references |
|---|---|---|---|
| Mechanistic | Upstream | DTP | mAb glycosylation: Developed a model linking extracellular nutrients to glycan patterns [39], and incorporated an in-silico reconstruction of sugar donor metabolism [40] |
| | | | mRNA production: First kinetic model incorporating magnesium pyrophosphate formation [43] and the effect of pyrophosphatase [44] |
| | | | rAAV production: Created mechanistic models for mammalian (transient triple transfection, [49], effect of Rep protein expression [51]) and insect cells [50] |
| | | | Viral vector transduction: Simulator for transduction in various culture modes [53] |
| Mechanistic | Downstream | Primarily DTP; [57] is DTI | Protein A Chromatography: Predicted loading/elution under varying conditions and resin types [63] |
| | | | Ion Exchange Chromatography: Modeled mAb retention factors (pH/salt) [61] and implemented a digital twin instance (online HPLC) for real-time optimization [57] |
| | | | Size-Exclusion Chromatography: Investigated adsorption of mAbs and impurities (aggregates/fragments) on a novel resin [64] |
| | | | Multimodal Chromatography: Model to predict mAb elution profile under different pH using protein sequence information [68] |

| Mechanistic | Freeze-drying | DTP | Primary drying: Modeled sublimation/heat transfer to optimize shelf temperature and chamber pressure [69, 70]<br><br>Thermal effects: Analyzed radiation on product uniformity [71]<br><br>Continuous freeze-drying [72] |
|---|---|---|---|
| Data-driven | Upstream | DTP | mAb Glycan Prediction: Built ANNs for predicting glycan distributions in CHO cells [78]<br><br>E. coli Fermentation: Used data-driven models for biomass/protein concentration [81] |
| Data-driven | Downstream | DTP | Purification (Fab products): Optimized resin selection, column sizing, and process steps to reduce overall cost [82] |
| Hybrid | Upstream | DTP | Fed-Batch mAb Production: Combined ANN/mass balance equations to predict key variables, validated on large datasets [86]<br><br>CHO Cell Culture: Integrated mass balance with ML (random forest, XGBoost) to predict viable cell density, titer, and glucose consumption [87] |

**Table 4: Current trends in PAT implementation for biopharmaceutical manufacturing. DTP = digital twin prototype, DTI = digital twin instance, DTA = digital twin aggregate**

| Study focus | Process stage | Parameters monitored | Analytical methods | Digital twin category | Key contributions | Reference |
|---|---|---|---|---|---|---|
| Protein concentration during mAbs production | Upstream | Protein stability under stress | Raman spectroscopy | Enabler of DTI | Identified protein stability under thermal and oxidative stress | [95] |
| Monitoring of mAb charge variants during cation exchange chromatography (CEX) | Downstream | Acidic, main, basic charge variants, total protein concentration | Raman spectroscopy, CNN | Enabler of DTI | CNN model was trained from spectra from mAb samples | [96] |
| Soft sensor for freeze-drying cycle | Lyophili-zation | Ice content, heat transfer coefficient, dried cake to vapor flow | Mechanistic model, Kalman filter | Enabler of DTI | Estimation of critical parameters during freeze-drying | [101] |

| Sof sensor for secondary during cycle | Lyophili-zation | Residual moisture | Mechanistic model | Enabler of DTI | Employed a state observer which uses temperature values, and mechanistic model of heat transfer and desorption kinetics | [102] |
| --- | --- | --- | --- | --- | --- | --- |
| Review of PAT technologies and data automation | General | Various CQAs and CPPs | Various spectroscopic methods | Enabler of DTI | Available PAT techniques, data automation, and visualization tools for real-time bioprocess monitoring | [94] |

**Table 5: Current trends in process control for biopharmaceutical manufacturing. DTP = digital twin prototype, DTI = digital twin instance, DTA = digital twin aggregate**

| Process stage | Control approach | Digital twin category | Key contributions and references |
|---|---|---|---|
| Upstream | MPC, feedforward | DTP; [125] is DTI | Fed-batch mAb manufacturing (MPC): Multivariable control system to simultaneously optimize titer and glycosylation [120]<br><br>Fed-batch cell culture: MPC design to optimize feeding to maximize daily protein production [121]<br><br>Continuous viral inactivation: Controlled pH/residence time [125] |
| Downstream | DCS, Feedback, feedforward | DTI | CHO clarification (DCS): Real-time control of turbidity and cell separation efficiency via acoustic power [123, 124]<br><br>Purification of VLPs: Strategy to control column switching timing based on product breakthrough [122]<br><br>Ion exchange: Model based feedforward control strategy for controlling charge variants in mAb purification [57] |
| Process-wide | SCADA | DTA | Continuous mAb manufacturing: Control of integrated unit operations using mechanistic models and real-time measurements [126] |

**Figures**

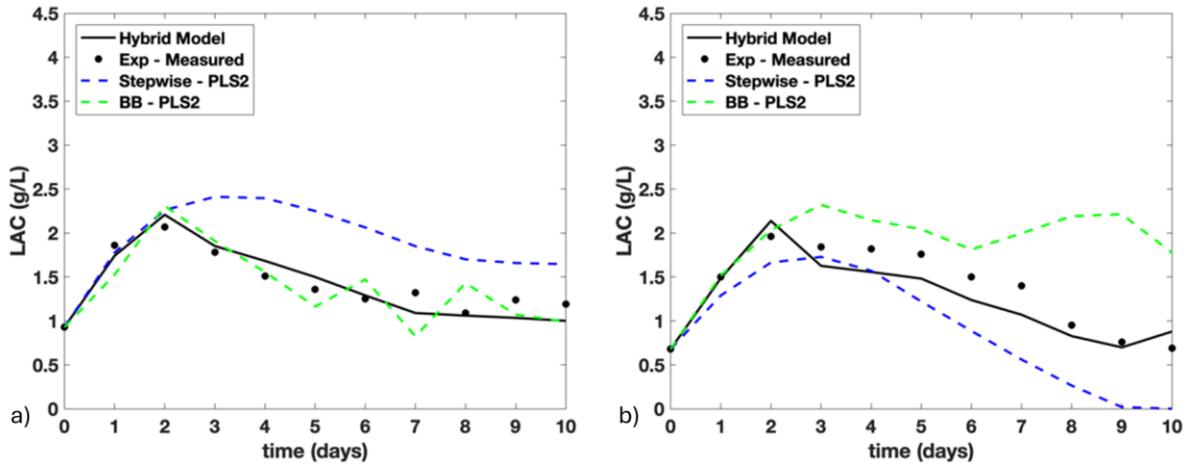

Fig. 1. Comparison of hybrid and data-driven model performance in predicting lactate concentration in two fed-batch runs of mAb production. Reprinted with permission from Narayanan et al. [86]. BB–PLS2 = black-box–partial least squares 2.

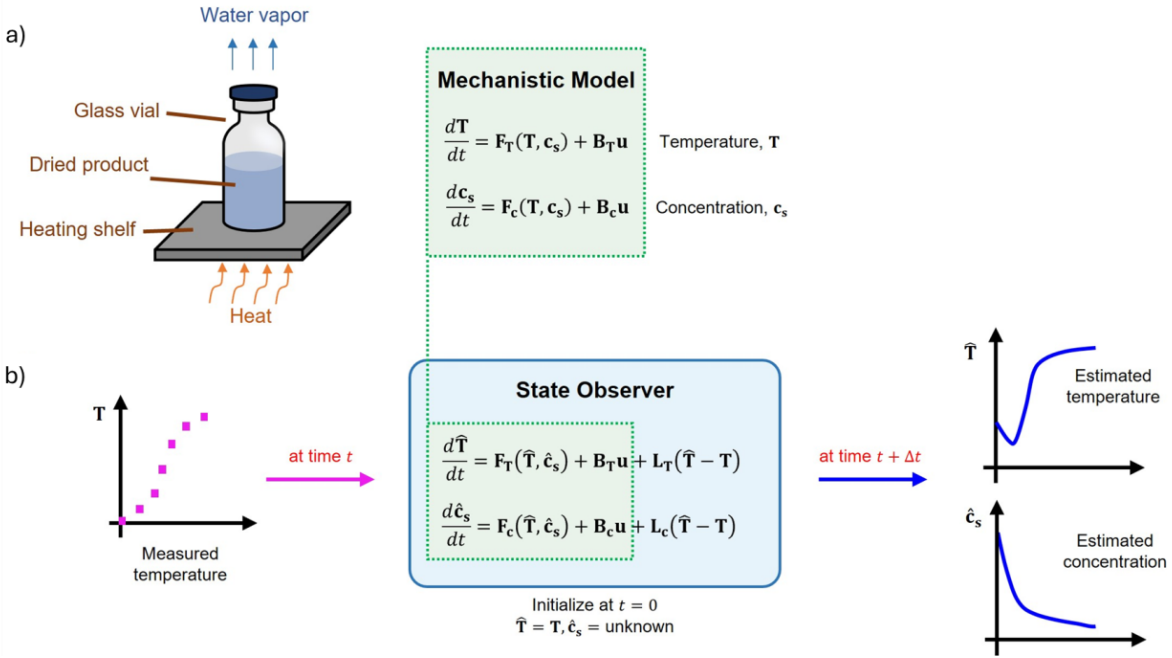

Fig. 2. a) Secondary drying in a lyophilization process. Mechanistic model describing the heat transfer and desorption dynamics. b) Overview of a state observer that receives measured temperature and control inputs (e.g., shelf temperature, microwave power) at each time step to estimate temperature and concentration for the next time step.

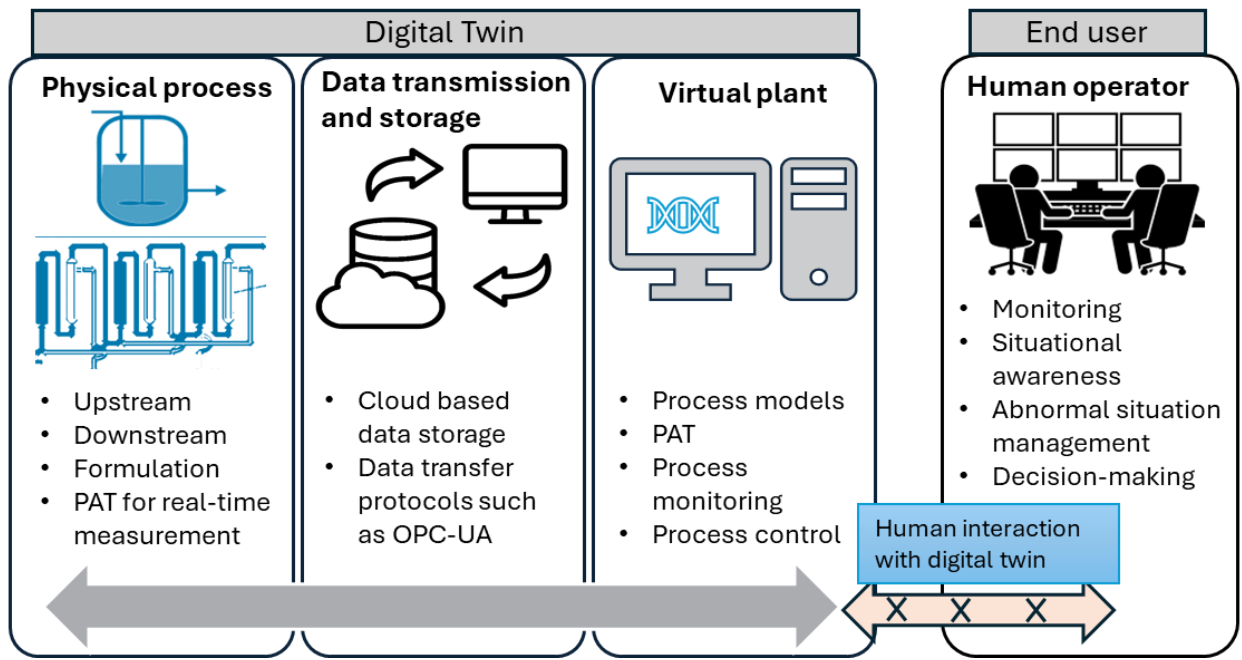

Fig. 3. Outline of the current state of digital twin implementation in biopharmaceutical manufacturing. The panel on the right illustrates that operator interaction is not adequately considered within the current framework.

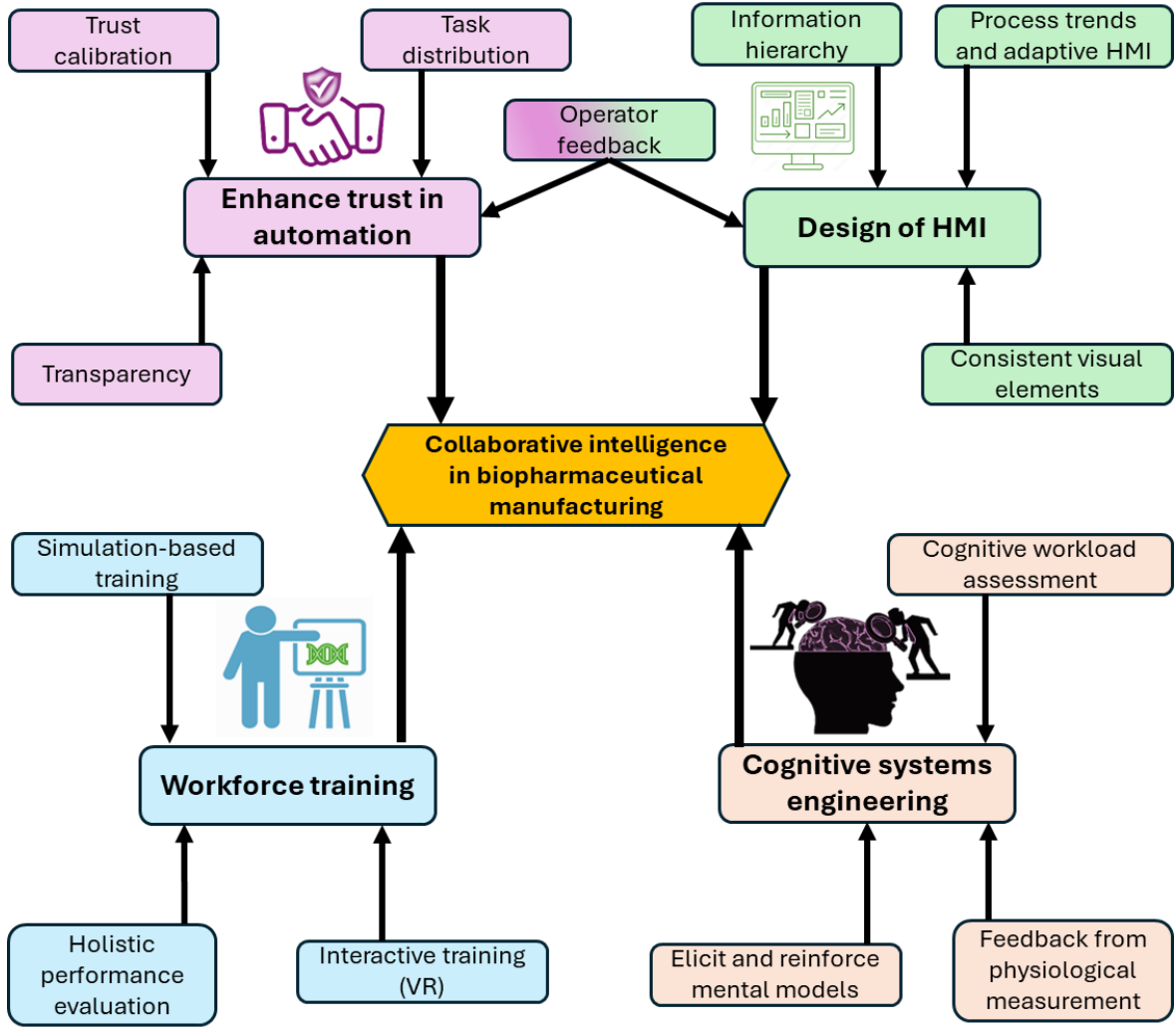

Fig. 4. Collaborative intelligence for digital twins in biopharmaceutical manufacturing. HMI = human-machine interface. VR = virtual reality.

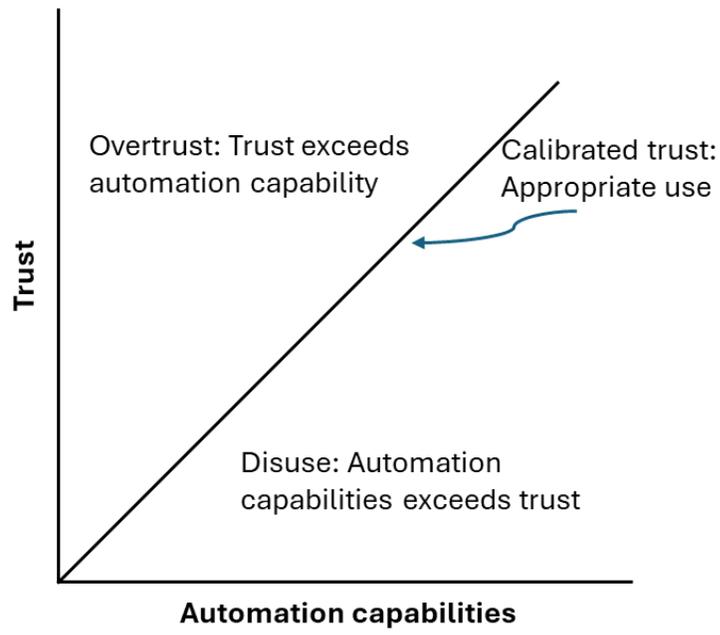

Fig. 5. Relationship between automation capabilities and operator trust [124].

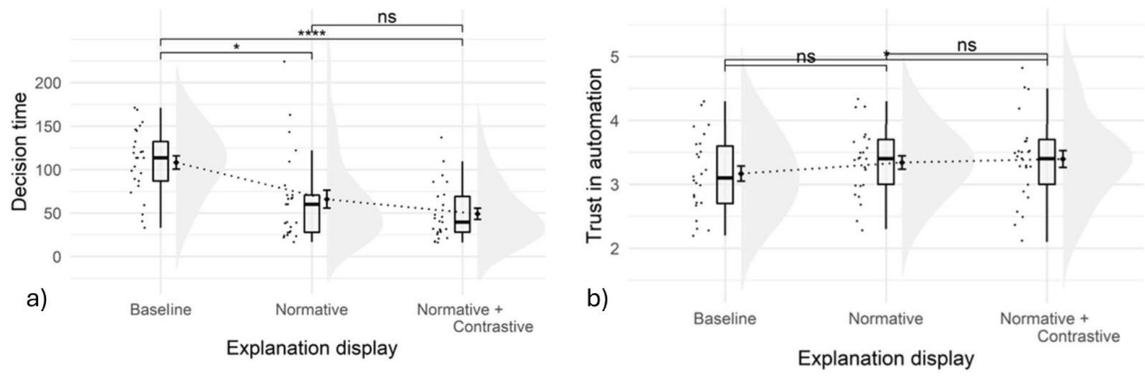

Fig. 6. Trends in a) decision time and b) trust in automation across three different interfaces with varying levels of explanation. Reprinted with permission from Gentile et al. [142].

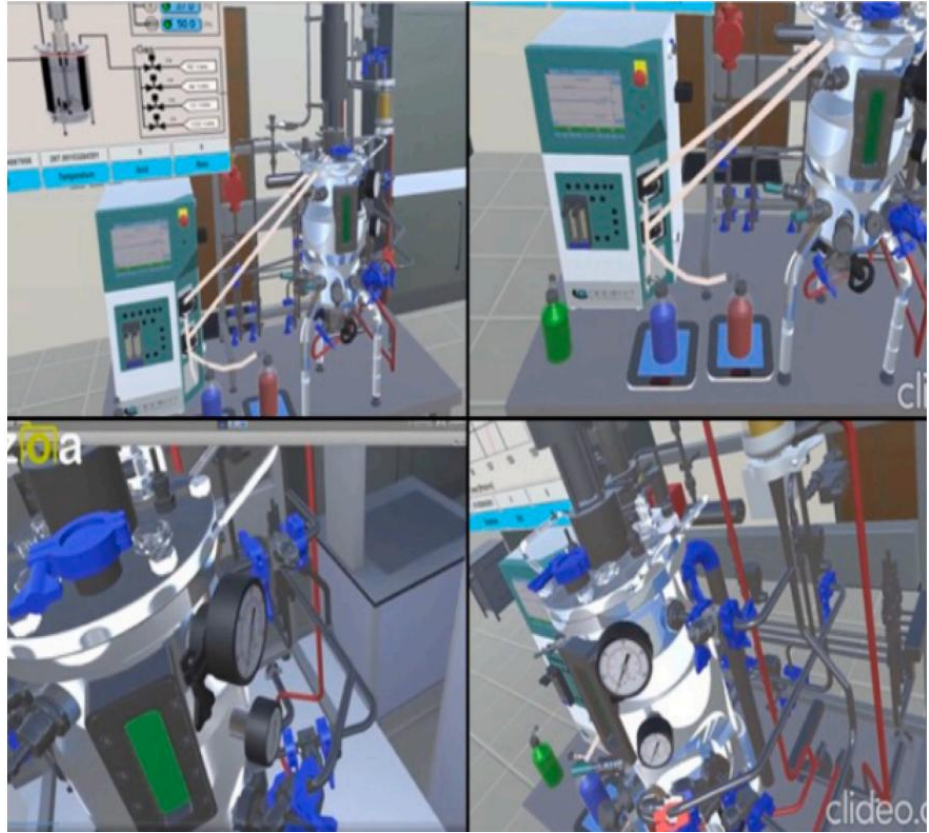

Fig. 7. Virtual-reality-based penicillin bioreactor simulator for operator training. Reprinted with permission from Hassan et al. [165].

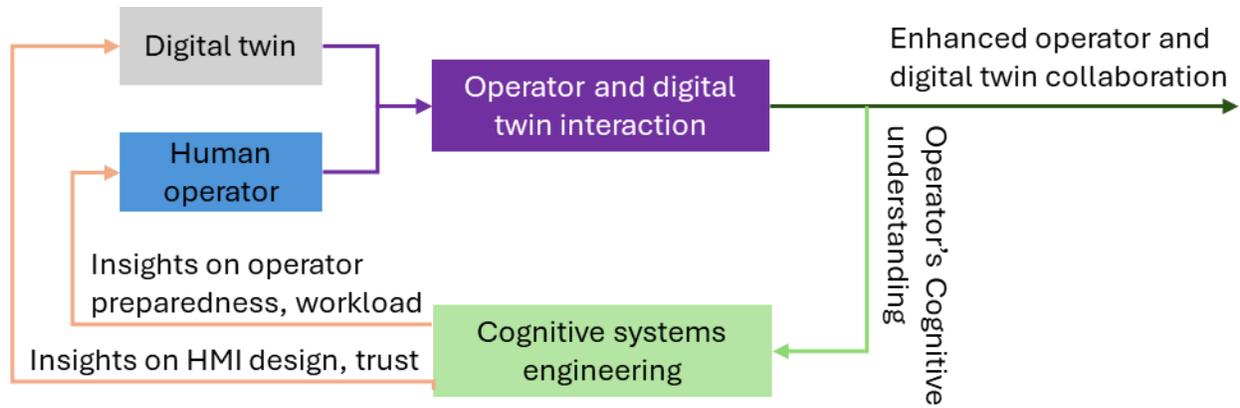

Fig. 8. Feedback loop to enhance collaborative intelligence in biopharma industries. HMI = human-machine interface.